\documentclass[final,11pt,onecolumn]{IEEEtran}

\usepackage{amssymb}
\usepackage{amstext}
\usepackage{latexsym}
\usepackage{times}
\usepackage{amssymb}
\usepackage{amsmath}
\usepackage{epsfig}
\usepackage{cite}
\usepackage{verbatim}
\usepackage{enumerate}
\usepackage{subfigure}
\usepackage{multicol}
\usepackage{pstricks}
\usepackage{pst-node}
\usepackage{setspace}
\usepackage{graphicx}
\usepackage{picinpar}
\usepackage{amsmath, amsfonts}
\usepackage{algorithm, algorithmic}
\usepackage{graphicx, comment}
\usepackage{stfloats}

\newtheorem{theorem}{Theorem}
\newtheorem{lemma}[theorem]{Lemma}
\newtheorem{remark}{Remark}

\begin{document}
\bibliographystyle{ieeetr}

\title{On Non-Integer Linear Degrees of Freedom of Constant Two-Cell MIMO Cellular Networks}

\author{
\authorblockN{Edin Zhang, Chiachi Huang, and Huai-Yan Feng}
\thanks{
E. Zhang and C. Huang are with the Department of Communications Engineering, Yuan Ze University, Taoyuan, Taiwan
(e-mail: s1014852@mail.yzu.edu.tw; chiachi@saturn.yzu.edu.tw).
H.-Y. Feng is with the Graduate Institute of Communication Engineering, National Taiwan University, Taipei, Taiwan
(e-mail: r02942118@ntu.edu.tw).
The material in this paper was presented in part at IEEE Information Theory Workshop, Hobart, Tasamania, Australia, 2014.
This work was supported by Ministry of Science and Technology, Taiwan, under Grants NSC-102-2221-E-155-012 and MOST-103-2221-E-155-011.}}

\maketitle

\begin{abstract}
The study of degrees of freedom (DoF) of multiuser channels has led to the development of important interference managing schemes, such as interference alignment (IA) and interference neutralization.
However, while the integer DoF have been widely studied in literatures, non-integer DoF are much less addressed, especially for channels with less variety.
In this paper, we study the non-integer DoF of the time-invariant multiple-input multiple-output (MIMO) interfering multiple access channel (IMAC) in the simple setting of two cells, $K$ users per cell, and $M$ antennas at all nodes.
We provide the exact characterization of the maximum achievable sum DoF under the constraint of using linear interference alignment (IA) scheme with symbol extension.
Our results indicate that the integer sum DoF characterization  $2MK/(K+1)$ achieved by the Suh-Ho-Tse scheme can be extended to the non-integer case only when $K \leq M^2$ for the circularly-symmetric-signaling systems and $K \leq 2M^2$ for the asymmetric-complex-signaling systems.
These results are further extended to the time-invariant parallel MIMO IMAC with independent subchannels.
\end{abstract}


\newpage

\section{Introduction}
Multiple-input multiple-output (MIMO) systems are capable of providing remarkably higher capacity compared to traditional single-input single-out (SISO) systems. The multiple antennas provide the extra dimensions to multiplex signals in space or cancel the interference from multiple unintended transmitters. The number of degrees of freedom (DoF), also known as multiplexing gain or capacity prelog, is a high signal-to-noise ratio (SNR) capacity approximation and characterizes the resolvable signal dimensions of the system.
Interference alignment (IA) is an interference managing scheme developed from the study of the degrees of freedom of the time-invariant two-transmitter MIMO $X$ channel \cite{Jafar_Shamai,MMK}. The IA scheme is shown to be DoF-optimal for the channel, and the concept of IA has been later applied to many fundamental channels, including the time-varying $K$-user SISO interference channel (IC) that provides $\frac{K}{2}$ sum DoF \cite{Cadambe_Jafar_int}, establishing the unbounded multiuser DoF gain of the channel. Also in \cite{Cadambe_Jafar_int}, Cadambe and Jafar develop a closed-form IA scheme to achieve $\frac{3M}{2}$ sum DoF for the time-invariant $3$-user MIMO IC with $M$ antennas at each node, and their DoF-optimal IA scheme can be implemented simply by linear precoder and combiner.

Interfering multiple access channel (IMAC) consists of several traditional multiple access channels (MAC) that interfere with each other, and IMAC is of practical importance because it models the environment of the uplink communications of several adjacent cells. Suh and Tse \cite{Suh_Tse} develop a linear IA scheme for the time-invariant two-cell MIMO IMAC, where there are $K$ users in each cell and all nodes are equipped with $M$ antennas, to achieve $\frac{2MK}{K+1}$ sum DoF, under the requirement that $M=K+1.$ The promising result indicates that the same DoF of the two isolated MACs, i.e., $2M$, can be realized when $K$ approaches infinity, demonstrating the multiuser DoF gain of the channel.

The application of the IA schemes to the two-cell IMAC has been extended in many directions, including the dual interfering broadcast channel (IBC) in \cite{Suh_Ho_Tse, Shin_Lee_Lim_Shin_Jang, Liu_Yang_Feasibility, Liu_Yang, Shi_Razaviyayn_Luo_He, Tang_Lambotharan}, the more general antenna settings in \cite{Shin_Lee_Lim_Shin_Jang, Liu_Yang_Feasibility, Liu_Yang, Shi_Razaviyayn_Luo_He, Tang_Lambotharan, Kim_Love_Clerckx_Hwang, Sridharan_Yu_Globecom, Ayoughi_Nasiri_Kenari_Khalaj, Chou_Chou_Wu_Chang}, and the more general $C$-cell settings in \cite{Liu_Yang_Feasibility, Liu_Yang, Shi_Razaviyayn_Luo_He, Tang_Lambotharan, Kim_Love_Clerckx_Hwang}.
Suh, Ho, and Tse study the dual time-invariant two-cell MIMO IBC in \cite{Suh_Ho_Tse}, and show that the same $\frac{2MK}{K+1}$ sum DoF can be achieved by linear IA scheme with less exchange of channel state information compared to that of the two-cell IMAC \cite{Suh_Tse}.
The time-invariant IMAC with $K$ users per cell, $M$ antennas at each transmitter, and $N$ antennas at each receiver, which will be referred to as the $(C,K,M,N)$ IMAC later in this paper, is studied by Kim et al in \cite{Kim_Love_Clerckx_Hwang}, and they provide an upper bound for the sum DoF of the channel.
Liu and Yang study the time-invariant $(C,K,M,N)$ IBC in \cite{Liu_Yang_Feasibility} and \cite{Liu_Yang}, where \cite{Liu_Yang_Feasibility} derives the feasibility condition of the linear IA scheme and \cite{Liu_Yang} obtains the characterization of the sum DoF. However, while significant progress has been made, the issue of the non-integer DoF is addressed neither in \cite{Liu_Yang_Feasibility} due to the assumption of no symbol extension to provide the generic channel matrices required by the algebraic structure nor in \cite{Liu_Yang} due to idea of the spatial extension \cite{Wang_Guo_Jafar} that avoids the DoF rounding.
Although a standard method to provide non-integer DoF is through symbol extension, the limitation of using symbol extension is in general still unknown and plays a key role in the study of non-integer DoF.

The non-integer DoF of time-invariant channels achieved by linear IA schemes with symbol extensions is considered in \cite{Wang_Guo_Jafar, Li_Jafarkhani_Jafar, Li_Jafarkhani}, and their results show that the block-diagonal structure of the symbol-extended channel matrices, where all blocks are the same, provides extra constraints on linear precoding and combining. More specifically, Li, Jafarkhani, and Jafar show that the number of independent variables in the symbol-extended channel matrix, which is termed channel diversity in \cite{Bresler_Tse_diversity}, limits the resolvability of the desired signal subspace and the interference subspace \cite{Li_Jafarkhani_Jafar}. Under the assumptions of linear processing, each transmitter sending the same number of data streams, and $M$ antennas at each node, DoF upper bounds for both the $X$ channel and the $K$-user IC are obtained in \cite{Li_Jafarkhani_Jafar} based on the channel diversity. However, achievability of their DoF upper bounds is not addressed and therefore still an open problem.
The non-integer DoF of the time-invariant $(C,K,M,N)$ IMAC is also still an open problem, except for the the special case of $(C,K,M,N)=(2,2,2,2)$ studied in \cite{Li_Jafarkhani}.

As a stepping stone to explore the non-integer DoF of the time-invariant $(C, K, M, N)$ IMAC, we study the model in the simple setting of two cells, $K$ users per cell, and $M$ antennas at all nodes. Since $M$ is fixed, no spatial extension \cite{Wang_Guo_Jafar} is allowed.
Moreover, all nodes are constrained to use linear pre- and post- processing due to the implementation issue, and symbol extension with \emph{arbitrary} numbers of time slots is allowed.
We apply the idea of channel diversity to this setting, and by exploring the block-diagonal structure of the symbol-extended channel matrices, we propose a novel upper bound and a modified lower bound for the sum DoF of the channel.
In particular, the converse is developed by deriving a rank ratio inequality, which is originally proposed for the time-varying $X$ channel \cite{Lashgari_Avestimehr_Suh}, for the the time-invariant, symbol extended IMAC. And the achievability is obtained by utilizing the generic structure imposed by the scheme design.
The tightness of the upper bounds is shown, and we obtain the exact characterization of the maximum linearly achievable sum DoF.
Our results indicate that the integer sum DoF characterization $\frac{2MK}{K+1}$ achieved by the linear IA scheme \cite{Suh_Tse, Suh_Ho_Tse} can be extended to the non-integer case only when $K \leq M^2$ for the traditional circularly-symmetric-signaling (CSS) systems and $K \leq 2M^2$ for the less-traditional asymmetric-complex-signaling (ACS) \cite{Cadambe_Jafar_Wang} systems due to the channel diversity constraint.
These results are further extended to the time-invariant parallel IMAC with independent subchannels.

The rest of the paper is organized as follows. Section \ref{sec:model} describes the models. Section \ref{sec:results} summarizes our main results. In Sections \ref{sec:proof_ccs} and \ref{sec:proof_acs}, we prove the theorems for the CSS and ACS systems, respectively. Section \ref{sec:parallel} extends the results to the parallel channels and Section \ref{sec:conclusion} concludes the paper.

Regarding notation usage, we use $\mathbf{O}_{m \times n}$, $\mathbf{I}_n$, $\mathbf{0}_n$, and $\mathbf{e}_m$ to respectively denote the $m \times n$ zero matrix, the $n \times n$ identity matrix, the $n \times 1$ zero vector, and the elementary column vector whose elements are all zero except that the $m^{th}$ element is 1. $\mathbf{A}^{-1}$, $\mathbf{A}^{t}$, $\mathbf{A}^{\dagger}$, and $\textrm{vec}(\mathbf{A})$ denote the inverse, the transpose, the conjugate transpose, and the vectorization operation of a matrix $\mathbf{A}$, respectively. We use $\textrm{blck}(\mathbf{A}_1, \ldots, \mathbf{A}_n)$ to denote the block-diagonal matrix with blocks $\mathbf{A}_1, \ldots, \mathbf{A}_n$, and $(\mathbf{x})_i$ to denote the $i^{th}$ element of a vector $\mathbf{x}$.

\section{System Model}
\label{sec:model}

\begin{figure}
\begin{center}
\includegraphics[width=180pt, trim=63 90 23 70,clip ]{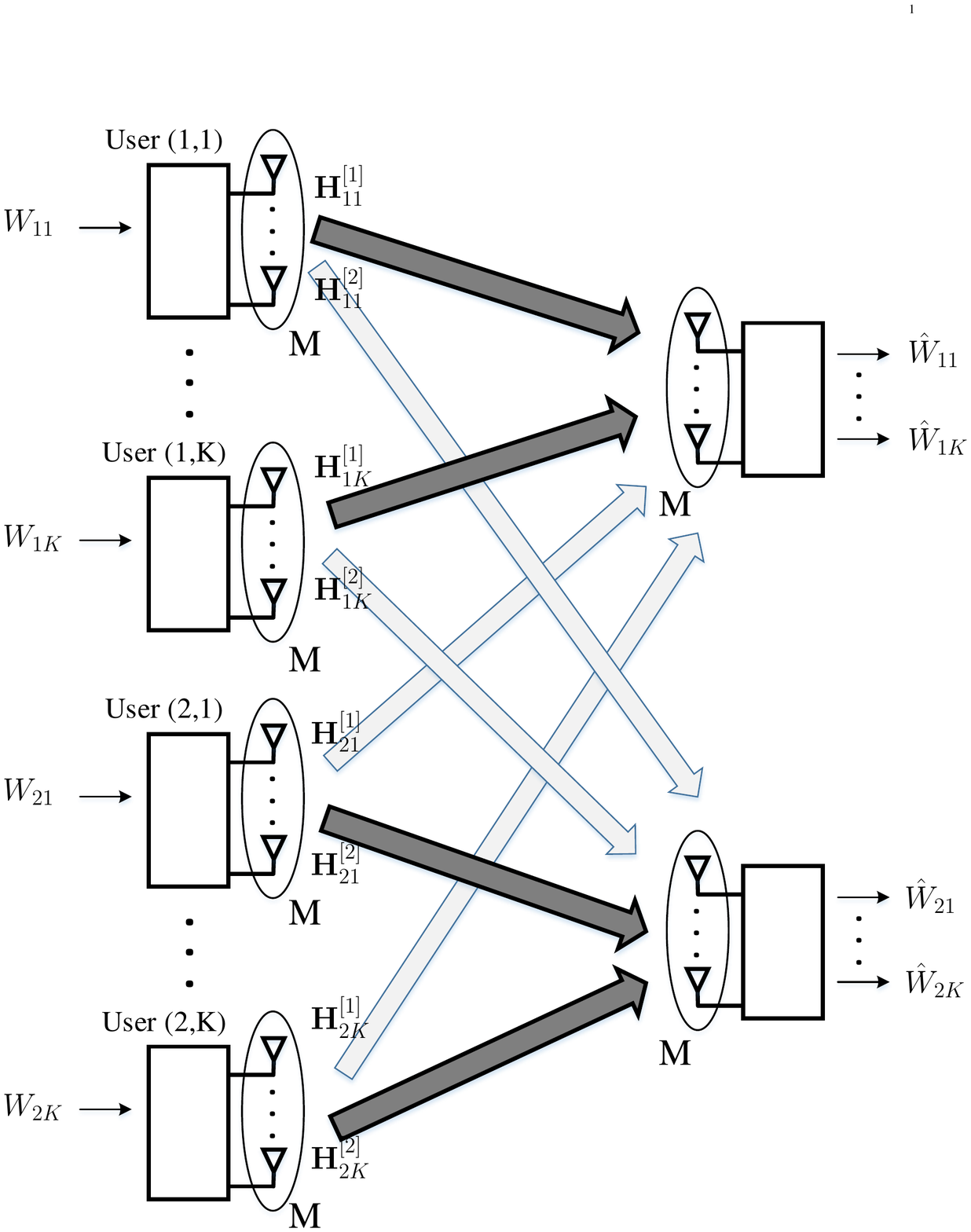}
\caption{The two-cell MIMO interfering multiple access channel.}
\label{fig:Systen_Model}
\end{center}
\end{figure}

Consider the time-invariant two-cell MIMO interfering multiple access channel with $K$ users in each cell and $M$ antennas at each node.
The channel is described by the input-output equation given as
\begin{align}
\label{eqn:channel}
\mathbf{y}^{[r]}(t)
=
\sum_{c=1}^2
\sum_{k=1}^K
\mathbf{H}_{ck}^{[r]}
\mathbf{x}_{ck}(t)
+
\mathbf{z}^{[r]}(t), ~~r=1,2
\end{align}
where at the $t^{th}$ channel use,
$\mathbf{y}^{[r]}(t)$,
$\mathbf{z}^{[r]}(t)$ are the $M \times 1$ vectors
representing the channel output and additive white Gaussian noise at receiver $r$,
$\mathbf{H}_{ck}^{[r]}$ is
the $M \times M$ channel matrix from transmitter $k$ in cell $c$ to receiver $r$,
and $\mathbf{x}_{ck}(t)$ is the
$M \times 1$ channel input from transmitter $k$ in cell $c$, for
$r, c \in \{ 1, 2 \}$ and $k \in \{ 1, \ldots, K \}$.
The elements of $\mathbf{H}_{ck}^{[r]}$ are assumed to be outcomes of independent and identically distributed (i.i.d.) continuous random variables and do not change with $t$. The elements of $\mathbf{z}^{[r]}(t)$, $r=1,2$, are
i.i.d. (both across space and time) circularly symmetric complex Gaussian random variables with zero mean and unit variance.
Following the existing works in literature, we assume that all channel matrices are known by all nodes in the channel. Note that the value of $M$ is part of the model description, and therefore the idea of spatial extension \cite{Wang_Guo_Jafar} by adding more antennas as an achievable scheme of the channel is not allowed.
The transmit power constraint is expressed as
\begin{align}
\label{eqn:power_cons}
\mbox{E}[|| \mathbf{x}_{ck}(t) ||^2]\leq P.
\end{align}

There are $2K$ independent messages $W_{ck}$, $c \in \{ 1, 2\}$ and $k \in \{ 1, \ldots, K \}$, associated with rates $R_{ck}$ to be communicated from transmitter $k$ in cell $c$ to receiver $c$.
The capacity region $\mathcal{C}(P)$ is the set of all rate tuples $(R_{11}, \ldots, R_{2K}) \in \mathbb{R}_+^{2K}$
for which the probability of error can be driven arbitrarily close to zero by using suitably long codewords.
The DoF region is defined as
\begin{align}
\mathcal{D}
& =
\Big\{ (d_{11}, \ldots, d_{2K}) \in \mathbb{R}_+^{2K}:
\exists
(R_{11}, \ldots, R_{2K})\in \mathcal{C}(P) \nonumber \\
& \quad \quad \quad \quad
\mbox{ s.t.} ~~ d_{ck}=\lim_{P\rightarrow\infty}\frac{R_{ck}(P)}{\log(P)}, ~~ (c,k) \in \mathcal{C}_{\textrm{cell}} \times \mathcal{K} \Big\} \nonumber
\end{align}
where $\mathcal{C}_{\textrm{cell}} = \{ 1, 2 \}$ and $\mathcal{K} = \{ 1, \ldots, K \}$.
The sum DoF is defined as
\begin{align}
d_{\textrm{it}}(K, M) = \max_{(d_{11}, \ldots, d_{2K}) \in \mathcal{D}}
d_{11} + \cdots + d_{2K}.
\end{align}
We include the indices $M$, $K$ to denote the $d_{\textrm{it}}$ for different $M$ and $K$.

In this paper, we study the sum DoF achieved by IA scheme in signal space with symbol extension, where an arbitrary number of time slots $T$ is allowed.
We consider both the traditional CSS system and the ACS system described respectively in the following two subsections.

\subsection{Circularly-Symmetric-Signaling System}
\label{subsec:CSS}
For a CSS system with $T$-symbol extension, the input-output relationship of the extended channel is given as
\begin{align}
\label{eqn:CCSchennel}
\bar{\mathbf{y}}^{[r]}
=
\sum_{c=1}^2
\sum_{k=1}^K
\bar{\mathbf{H}}_{ck}^{[r]}
\bar{\mathbf{x}}_{ck}
+
\bar{\mathbf{z}}^{[r]}, ~~r=1,2
\end{align}
where the $MT \times MT$ matrix $\bar{\mathbf{H}}_{ck}^{[r]}$ is given as
\begin{align}
\label{eqn:extended_H}
\bar{\mathbf{H}}_{ck}^{[r]}
=
\textrm{blck}(\mathbf{H}_{ck}^{[r]}, \ldots, \mathbf{H}_{ck}^{[r]})
\end{align}
and
\begin{align}
\label{eqn:scheme_IMAC}
\bar{\mathbf{x}}_{ck} =
\left[
\begin{array}{c}
\mathbf{x}_{ck}(1) \\
\vdots \\
\mathbf{x}_{ck}(T)
\end{array}
\right]
\in
\mathbb{C}^{MT}.
\end{align}
Similar notation applies to $\bar{\mathbf{y}}^{[r]}$ and $\bar{\mathbf{z}}^{[r]}$.
Linear precoding and combining in extended signal space are described as
\begin{align}
\bar{\mathbf{x}}_{ck} =
\bar{\mathbf{V}}_{ck}
\mathbf{s}_{\textrm{css},ck}, ~~~~
\hat{\mathbf{s}}_{\textrm{css},ck} =
\bar{\mathbf{U}}_{ck}^{\dagger}
\bar{\mathbf{y}}^{[c]}
\end{align}
where
$\bar{\mathbf{V}}_{ck} \in \mathbb{C}^{MT \times n_{ck}}$ is the precoding matrix and
$\bar{\mathbf{U}}_{ck} \in \mathbb{C}^{MT \times n_{ck}}$ is the combining matrix for user $k$ in cell $c$ that sends $n_{ck} \in \mathbb{Z}_+$ data streams described by the source vector $\mathbf{s}_{\textrm{css},ck} \in \mathbb{C}^{n_{ck} \times 1}$.
For simplicity, we assume that the data streams are independent by requiring
$\mathrm{E} [
\mathbf{s}_{\textrm{css},ck}
\mathbf{s}_{\textrm{css},ck}^{\dagger}] = \mathbf{I}_{n_{ck}}$.
To introduce the flexibility of a transmitter not to send any information, which is allowed in both the practical operation and the theoretic analysis of the capacity and DoF regions, to the feasibility analysis, we let $\bar{\mathbf{x}}_{ck} =
\bar{\mathbf{V}}_{ck} = \bar{\mathbf{U}}_{ck} = \mathbf{0}_{MT}$ when $n_{ck} = 0$.
The feasibility condition of the linear IA scheme \cite{Liu_Yang_Feasibility} in the $T$-symbol extended signal space is
\begin{align}
\label{eqn:align_condition1}
\mathrm{rank} \left(
\bar{\mathbf{U}}_{ck}^{\dagger}
\bar{\mathbf{H}}_{ck}^{[c]}
\bar{\mathbf{V}}_{ck}
\right)
&= n_{ck}  \\
\label{eqn:align_condition2}
\bar{\mathbf{U}}_{c'k'}^{\dagger}
\bar{\mathbf{H}}_{ck}^{[c']}
\bar{\mathbf{V}}_{ck}
&= \mathbf{O}, ~~~~~\textrm{if} ~ (c', k') \neq (c, k)
\end{align}
for all
$c, c' \in \{ 1, 2 \}$ and $k, k' \in \{ 1, \ldots, K \}$.
We further define the feasible sum DoF $d_{\textrm{f},\textrm{css}} \in \mathbb{Q_+}$ that represents the largest number of sum degrees of freedom achieved by linear IA scheme in signal space with finite symbol extension for the CSS system as
\begin{align}
\label{eqn:d_f_css_def}
d_{\textrm{f},\textrm{css}}(K, M) = \max_{T \in \mathbb{Z}_+}
\left\{
\frac{1}{T}
\max_{(n_{11}, \ldots, n_{2K}) \in \bar{\mathcal{F}}_{T}}
n_{11} + \cdots + n_{2K}
\right\}
\end{align}
where
$\bar{\mathcal{F}}_T$
is the set of all $(n_{11}, \ldots, n_{2K}) \in \mathbb{Z}_+^{2K}$ satisfying
the IA condition (\ref{eqn:align_condition1}), (\ref{eqn:align_condition2}) with $T$-symbol extension.
We include the indices $M$, $K$ to denote the $d_{\textrm{f},\textrm{css}}$ for different $M$ and $K$.
Note that these definitions imply that $d_{\textrm{f},\textrm{css}} \leq d_{\textrm{it}}$.

\subsection{Asymmetric-Complex-Signaling System}
\label{subsec:ACS}
The main idea of ACS is to separate the real and imaginary parts of the transmit and receive signals \cite{Wang_Guo_Jafar}.
The input-output relationship of the extended channel for an ACS system with $T$-symbol extension is given as
\begin{align}
\label{eqn:ACSchennel}
\widetilde{\mathbf{y}}^{[r]}
=
\sum_{c=1}^2
\sum_{k=1}^K
\widetilde{\mathbf{H}}_{ck}^{[r]}
\widetilde{\mathbf{x}}_{ck}
+
\widetilde{\mathbf{z}}^{[r]}, ~~r=1,2
\end{align}
where
$\widetilde{\mathbf{H}}_{ck}^{[r]} \in \mathbb{R}^{2MT \times 2MT}$
is given as
\begin{align}
\label{eqn:extended_H_acs}
\widetilde{\mathbf{H}}_{ck}^{[r]}
=
\textrm{blck}(\check{\mathbf{H}}_{ck}^{[r]}, \ldots, \check{\mathbf{H}}_{ck}^{[r]})
\end{align}
where
\begin{align}
\label{eqn:extended_H_acs_structure}
\check{\mathbf{H}}_{ck}^{[r]}
=
\left[
\begin{array}{ccccc}
\textrm{Re}(h_{11}^{rck}) & -\textrm{Im}(h_{11}^{rck})    & \cdots & \textrm{Re}(h_{1M}^{rck}) & -\textrm{Im}(h_{1M}^{rck}) \\
\textrm{Im}(h_{11}^{rck}) & \textrm{Re}(h_{11}^{rck})     & \cdots & \textrm{Im}(h_{1M}^{rck}) & \textrm{Re}(h_{1M}^{rck}) \\
\vdots & \vdots & \ddots & \vdots & \vdots \\
\textrm{Re}(h_{M1}^{rck}) & -\textrm{Im}(h_{M1}^{rck})    & \cdots & \textrm{Re}(h_{MM}^{rck}) & -\textrm{Im}(h_{MM}^{rck}) \\
\textrm{Im}(h_{M1}^{rck}) & \textrm{Re}(h_{M1}^{rck})     & \cdots & \textrm{Im}(h_{MM}^{rck}) & \textrm{Re}(h_{MM}^{rck})
\end{array}
\right]
\end{align}
where $h_{ij}^{rck}$ is the $(i,j)$ element of $\mathbf{H}_{ck}^{[r]}$, and $\widetilde{\mathbf{x}}_{ck}$ is given as
\begin{align}
\label{eqn:tilde_x}
\widetilde{\mathbf{x}}_{ck} =
\left[
\begin{array}{c}
\check{\mathbf{x}}_{ck}(1) \\
\vdots \\
\check{\mathbf{x}}_{ck}(T)
\end{array}
\right]
\end{align}
where
\begin{align}
\label{eqn:check_x}
\check{\mathbf{x}}_{ck}(t) =
\left[
\begin{array}{c}
\textrm{Re}((\mathbf{x}_{ck}(t))_1) \\
\textrm{Im}((\mathbf{x}_{ck}(t))_1) \\
\vdots \\
\textrm{Re}((\mathbf{x}_{ck}(t))_M) \\
\textrm{Im}((\mathbf{x}_{ck}(t))_M)
\end{array}
\right]
\in
\mathbb{R}^{2M}.
\end{align}
Similar notation applies to $\widetilde{\mathbf{y}}^{[r]}$ and $\widetilde{\mathbf{z}}^{[r]}$.
To introduce the notation, linear precoding and combining for ACS systems, which are similar to those for CSS systems, are given as follows.
\begin{align}
\widetilde{\mathbf{x}}_{ck} =
\widetilde{\mathbf{V}}_{ck}
\mathbf{s}_{\textrm{acs},ck}, ~~~~
\hat{\mathbf{s}}_{\textrm{acs},ck} =
\widetilde{\mathbf{U}}_{ck}^{\dagger}
\widetilde{\mathbf{y}}^{[c]}
\end{align}
where
$\widetilde{\mathbf{V}}_{ck} \in \mathbb{R}^{2MT \times n_{ck}}$,
$\widetilde{\mathbf{U}}_{ck} \in \mathbb{R}^{2MT \times n_{ck}}$,
and $\mathbf{s}_{\textrm{acs},ck} \in \mathbb{R}^{n_{ck} \times 1}$, whose detail descriptions, along with their feasibility condition of IA scheme, are omitted for brevity.

The feasible sum DoF $d_{\textrm{f},\textrm{acs}} \in \mathbb{Q}_+$ for the ACS system is defined as
\begin{align}
d_{\textrm{f},\textrm{acs}}(K, M) = \max_{T \in \mathbb{Z}_+}
\left\{
\frac{1}{2T}
\max_{(n_{11}, \ldots, n_{2K}) \in \widetilde{\mathcal{F}}_{T}}
n_{11} + \cdots + n_{2K}
\right\}
\end{align}
where
$\widetilde{\mathcal{F}}_T$
is the set of all $(n_{11}, \ldots, n_{2K}) \in \mathbb{Z}_+^{2K}$ satisfying the IA condition with $T$-symbol extension, and where we use coefficient $\frac{1}{2T}$, instead of $\frac{1}{T}$ in (\ref{eqn:d_f_css_def}), because of the fact that one real data stream only provides $\frac{1}{2}$ DoF.
Note that these definitions imply that
$d_{\textrm{f},\textrm{css}} \leq d_{\textrm{f},\textrm{acs}} \leq d_{\textrm{it}}$.

\section{Main Results}
\label{sec:results}
We present our main results in this section. For the ease of comparison, we first summarize the important result in the literature as follows.
The sum DoF $d_{\textrm{it}}$ and the feasible sum DoF $d_{\textrm{f},\textrm{css}}$ of the time-invariant two-cell $K$-user IMAC with $M$ antennas at each node, as defined in Section \ref{sec:model}, satisfy
\begin{align}
\label{eqn:d_bounds_ccs}
2K \Biggl\lfloor \frac{M}{K+1} \Biggr\rfloor \leq d_{\textrm{f},\textrm{css}}(K,M) \leq d_{\textrm{it}}(K, M) \leq \frac{2KM}{K+1},
\end{align}
which is obtained by combining the lower bound from  \cite{Suh_Tse, Suh_Ho_Tse}, where CSS systems are considered, and the upper bound from \cite{Kim_Love_Clerckx_Hwang}.
This result can be easily extended to ACS systems by combining the ACS scheme with the IA scheme in \cite{Suh_Tse, Suh_Ho_Tse}, and the extended result is
\begin{align}
\label{eqn:d_bounds_acs}
K \Biggl\lfloor \frac{2M}{K+1} \Biggr\rfloor \leq d_{\textrm{f},\textrm{acs}}(K,M) \leq d_{\textrm{it}}(K, M) \leq \frac{2KM}{K+1}.
\end{align}
Note that in (\ref{eqn:d_bounds_ccs}) and (\ref{eqn:d_bounds_acs}), when $\frac{M}{K+1}$ and $\frac{2M}{K+1}$ are integers, the lower bounds meet the upper bound. Otherwise, the upper bound is not tight due to the floor operations.

Our main results are the exact characterizations of $d_{\textrm{f},\textrm{css}}$ and $d_{\textrm{f},\textrm{acs}}$ provided in the following theorems, whose proofs are deferred in Sections \ref{sec:proof_ccs} and \ref{sec:proof_acs}, respectively.

\begin{theorem}
\label{thm:d_f_ccs}
\begin{align}
\label{eqn:d_f_ccs}
d_{\textrm{f},\textrm{css}}(K, M) =
\left\{
\begin{array}{ll}
2KM/(K+1) & \mathrm{if} ~ K \leq M^2 \\
2M^3/(M^2 + 1) & \mathrm{if} ~ K > M^2.
\end{array}
\right.
\end{align}
\end{theorem}

\begin{theorem}
\label{thm:d_f_acs}
\begin{align}
\label{eqn:d_f_acs}
d_{\textrm{f},\textrm{acs}}(K, M) =
\left\{
\begin{array}{ll}
2KM/(K+1) & \mathrm{if} ~ K \leq 2M^2 \\
4M^3/(2M^2 + 1) & \mathrm{if} ~ K > 2M^2.
\end{array}
\right.
\end{align}
\end{theorem}

\begin{figure*}[t]
\begin{center}
\mbox{
\subfigure[]{\includegraphics[width=250pt, trim=110 272 125 270, clip]{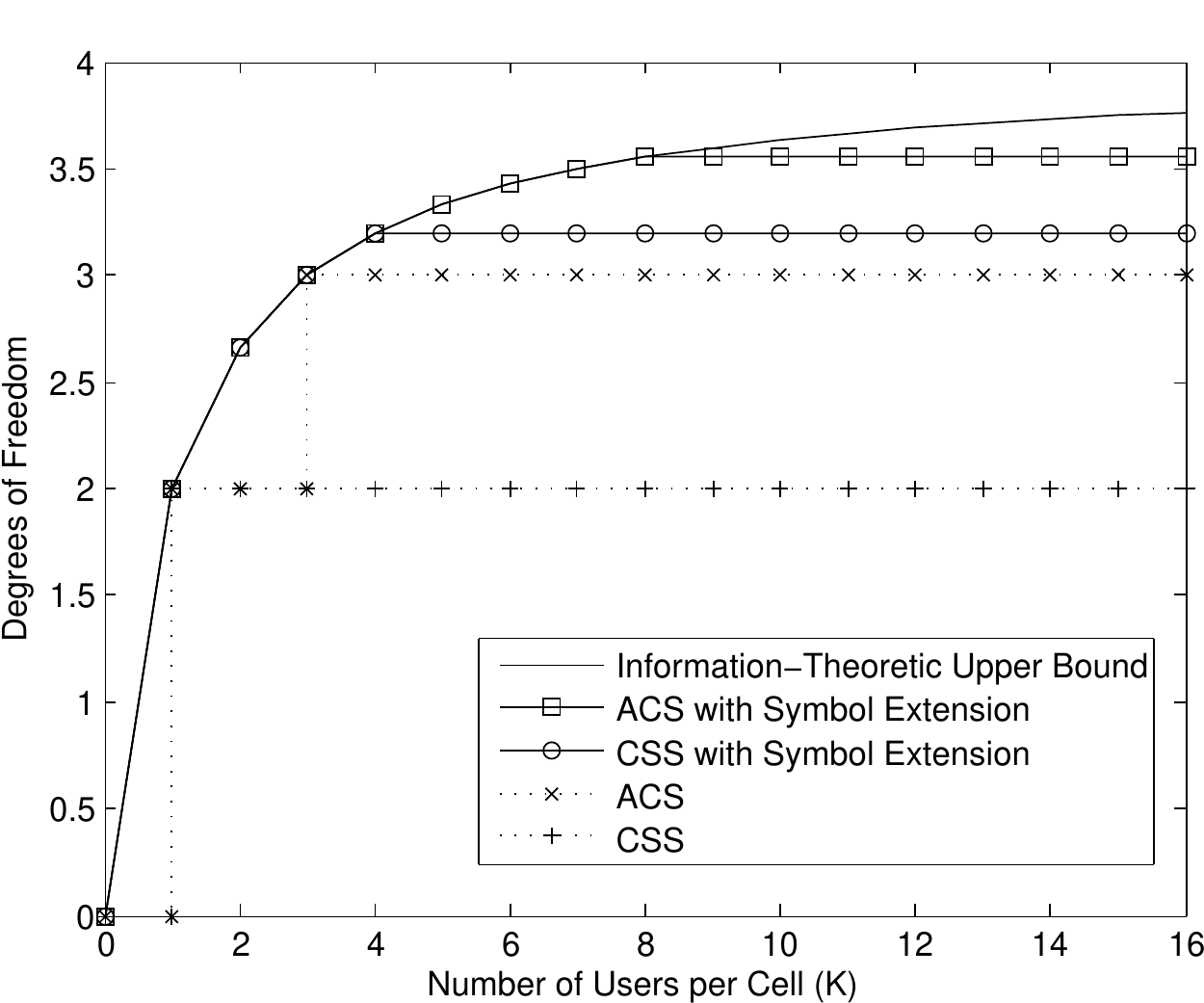}}
\subfigure[]{\includegraphics[width=250pt, trim=110 272 125 270, clip]{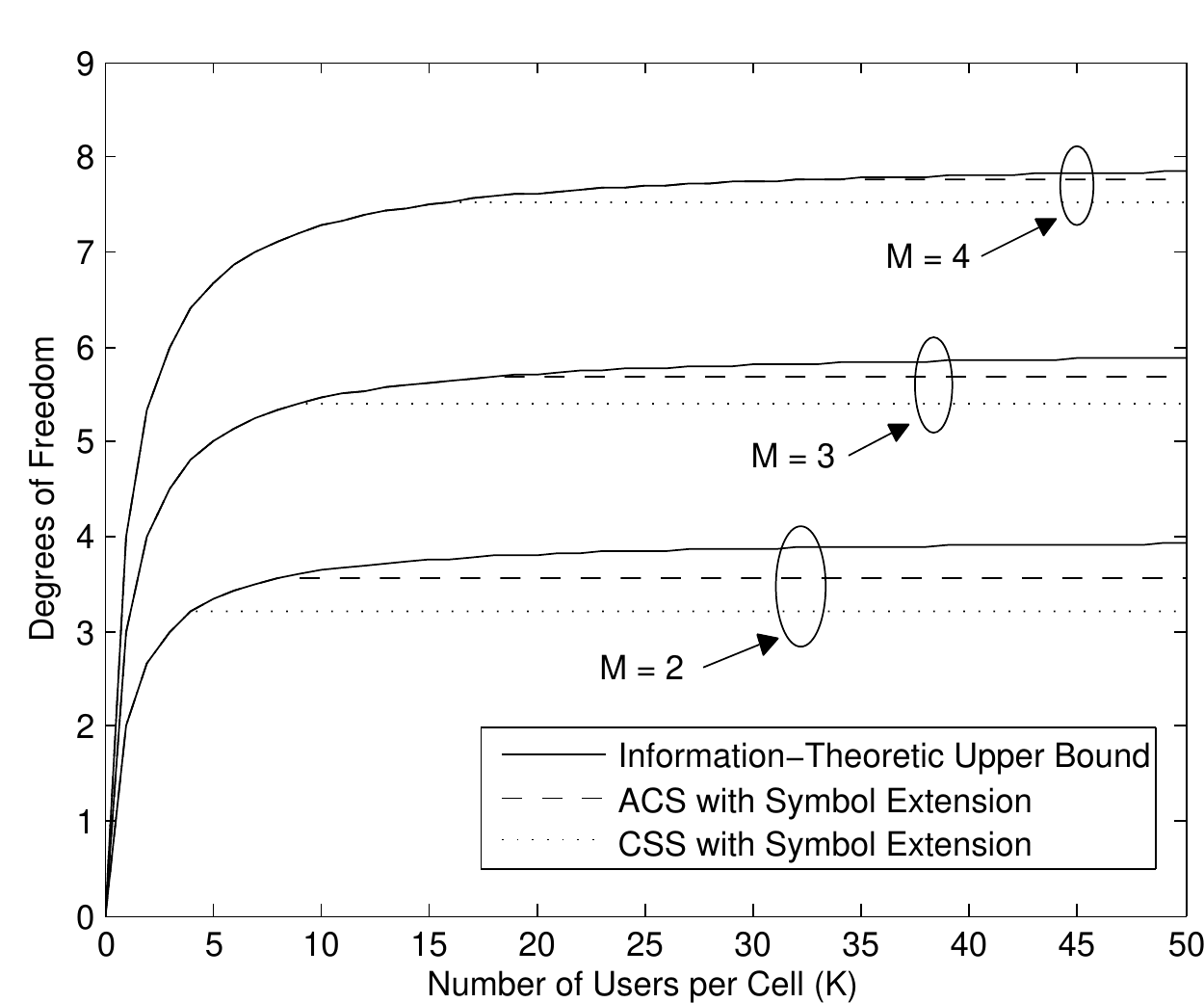}}
}
\end{center}
\caption{The largest numbers of sum DoF achieved by linear IA with symbol extension for CSS and ACS systems versus the number of users $K$ per cell.
The 2-antenna case ($M=2$) is plotted in (a), and 2-, 3-, and 4-antenna cases ($M=2,3,4$) are plotted in (b).
For comparison, the achievable integer DoF without using symbol extension $2K \left \lfloor \frac{M}{K+1} \right \rfloor$ given in \cite{Suh_Tse, Suh_Ho_Tse} for CSS system and $K \left \lfloor \frac{2M}{K+1} \right \rfloor$ for ACS system are also included in (a), and the information-theoretic upper bound \cite{Kim_Love_Clerckx_Hwang} is included in both (a) and (b).}
\label{fig:main_thm}
\end{figure*}

Our main results are illustrated in Fig. \ref{fig:main_thm}. We provide the following remarks on Theorems \ref{thm:d_f_ccs} and \ref{thm:d_f_acs}.
\begin{remark}
The loss of the achievable DoF caused by the floor operations in (\ref{eqn:d_bounds_ccs}) and (\ref{eqn:d_bounds_acs}) are removed in (\ref{eqn:d_f_ccs}) and (\ref{eqn:d_f_acs}) due to the symbol extension that helps provide non-integer DoF $\frac{M}{K+1}$ for each user.
\end{remark}

\begin{remark}
There are two different regimes of $K$ for $d_{\textrm{f},\textrm{css}}(K,M)$. When $K \leq M^2$, $d_{\textrm{f},\textrm{css}}$ increases as $K$ increases.
However, when $K > M^2$, $\frac{2MK}{K+1}$ is not feasible and the multiuser DoF gain disappears when using CSS linear IA scheme.
Similar observation can be made for $d_{\textrm{f},\textrm{acs}}(K,M)$.
These observations are summarized as
\begin{align}
\max_{K \in \mathbb{Z}_+} d_{\textrm{f},\textrm{css}}(K, M)
&=
2M(1-\frac{1}{M^2+1}) \\
\max_{K \in \mathbb{Z}_+} d_{\textrm{f},\textrm{acs}}(K, M)
&=
2M(1-\frac{1}{2M^2+1})
\end{align}
where $\frac{1}{M^2+1}$ and $\frac{1}{2M^2+1}$ represent the degrading factors of the two-cell interfering MAC from the two isolated MACs for CSS systems and ACS systems, respectively.
\end{remark}

\begin{remark}
With a slight notation abuse, we can combine the expressions for $d_{\textrm{f},\textrm{css}}$ and $d_{\textrm{f},\textrm{acs}}$ as follows. Let $d_{\textrm{f}}$ be the feasible sum DoF that includes $d_{\textrm{f},\textrm{css}}$ and $d_{\textrm{f},\textrm{acs}}$, understood by context. Then we can combine (\ref{eqn:d_f_ccs}) and (\ref{eqn:d_f_acs}) as
\begin{align}
d_{\textrm{f}}(K,M,D) = 2M \cdot \frac{K_{\textrm{act}}}{K_{\textrm{act}} + 1}
\end{align}
where $K_{\textrm{act}} = \min(K,D)$, which as explained later in Sections \ref{sec:proof_ccs} and \ref{sec:proof_acs} represents the number of active users in each cell, and $D$ is the channel diversity, which is $M^2$ for CSS systems and $2M^2$ for ACS systems. Now we can clearly see how $D$ translates into $K_{\textrm{act}}$, which in turn translates into $d_{\textrm{f}}$.
\end{remark}

\begin{remark}
Comparing $d_{\textrm{f},\textrm{css}}$, $d_{\textrm{f},\textrm{acs}}$, and $d_{\textrm{it}}$, we can divide parameters $K,M$ into three different regimes as follows.
For the first regime, where $K \leq M^2$, both CSS and ACS linear IA schemes with finite symbol extension achieve the DoF upper bound of the channel, i.e.,
\begin{align}
d_{\textrm{f},\textrm{css}} = d_{\textrm{f},\textrm{acs}} = d_{\textrm{it}} = \frac{2KM}{K+1}.
\end{align}
For the second regime, where $M^2 < K \leq 2M^2$, only
ACS linear IA scheme with finite symbol extension achieves the DoF upper bound, i.e.,
\begin{align}
d_{\textrm{f},\textrm{css}} < d_{\textrm{f},\textrm{acs}} = d_{\textrm{it}} = \frac{2KM}{K+1}.
\end{align}
For the last regime, where $K > 2M^2$, the characterization of $d_{\textrm{it}}$ and whether or not ACS linear IA scheme with finite symbol extension achieves the information-theoretic DoF are both still open problems for the considered time-invariant two-cell MIMO IMAC. However, we would like to mention that, on the contrary, for the time-varying setting, where the channel diversity constraint does not hold,
the characterization of the sum DoF for all $K,M$ can be shown to be $d_{\textrm{f},\textrm{css}} = d_{\textrm{f},\textrm{acs}} = d_{\textrm{it}} = \frac{2KM}{K+1}$ achieved by the CSS scheme given in \cite{Suh_Tse, Suh_Ho_Tse} with symbol extension.
\end{remark}

\section{Proof for CSS Systems}
\label{sec:proof_ccs}
In this section, we prove Theorem \ref{thm:d_f_ccs}, whose achievability and converse are stated separately in the following two theorems.
\begin{theorem}
\label{thm:d_f_low}
(\emph{Achievability:})
If $K \leq M^2$, then
\begin{align}
\label{eqn:d_f_low}
d_{\textrm{f},\textrm{css}}(K,M) \geq \frac{2KM}{K+1}.
\end{align}
\end{theorem}
\qquad \emph{Proof:}
The proof is given in Section \ref{sec:proof_ccs_low}.
\hfill $\blacksquare$

\vspace{2pt}

\begin{theorem}
\label{thm:d_f_up}
\emph{(Converse:)}
The feasible sum DoF satisfies
\begin{align}
\label{eqn:d_f_up}
d_{\textrm{f},\textrm{css}}(K,M) \leq \frac{2M^3}{M^2+1}.
\end{align}
\end{theorem}
\qquad \emph{Proof:}
The proof is deferred in Section \ref{sec:proof_ccs_up}.
\hfill $\blacksquare$

\vspace{2pt}

Theorem \ref{thm:d_f_ccs} is then proved by combining Theorem \ref{thm:d_f_low}, Theorem \ref{thm:d_f_up}, and the upper bound in (\ref{eqn:d_bounds_ccs}), and by communicating only to $M^2$ users in each cell when $K>M^2$ by letting some $\bar{\mathbf{V}}_{ck}$ be zero matrices. Now we proceed to prove the achievability and converse.

\subsection{Proof of Achievability}
\label{sec:proof_ccs_low}
We provide a constructive proof that focuses on the unsolved case, where $\frac{M}{K+1}$ is not an integer. The scheme is described as follows. The first step is to choose $T = K+1$, resulting in an extended signal space of $\mathbb{C}^{M(K+1)}$. The second step is to choose the $2M$ reference vectors
$\bar{\mathbf{r}}^{[1]}_1, \ldots, \bar{\mathbf{r}}^{[1]}_M, \bar{\mathbf{r}}^{[2]}_1, \ldots, \bar{\mathbf{r}}^{[2]}_M \in \mathbb{C}^{M(K+1)}$
by letting the elements of the vectors be outcomes of independent continuous random variables. The third step is to choose the precoding vectors.
Let the $m^{th}$ precoding vector of user $k$ in cell $c$ be
\begin{align}
\label{eqn:v_design1}
\bar{\mathbf{v}}_{ck,m} &=
\frac{\sqrt{(K+1)P}}{\sqrt{M} \big|\big| \bar{\mathbf{H}}_{ck}^{[\bar{c}]^{-1}}
\bar{\mathbf{r}}^{[\bar{c}]}_m \big|\big|}
\bar{\mathbf{H}}_{ck}^{[\bar{c}]^{-1}}
\bar{\mathbf{r}}^{[\bar{c}]}_m
\end{align}
for $m=1,\ldots,M$ and $(c, \bar{c}) = (1, 2), (2, 1)$. Note that the construction ensures that all $\bar{\mathbf{v}}_{1k,m}$, $k=1,\ldots,K$ align on
$\bar{\mathbf{r}}^{[2]}_m$ in the extended receive signal space at receiver 2.
The fourth step is to choose the combining vectors.
Let the $m^{th}$ combining vector of user $k$ in cell $c$ be
\begin{align}
\bar{\mathbf{u}}_{ck,m} &=
\mathrm{null}
\left(
\mathcal{R}_c
\cup
\left\{
\bar{\mathbf{H}}^{[c]}_{ck'} \bar{\mathbf{v}}_{ck',m'}:
(k',m') \neq (k,m)
\right\}
\right) \nonumber
\end{align}
where
$\mathcal{R}_c = \{ \bar{\mathbf{r}}^{[c]}_1, \ldots, \bar{\mathbf{r}}^{[c]}_M \}$ and for $c=1,~2$.
The last step is to construct the precoding matrix and combining matrix as
\begin{align}
\bar{\mathbf{V}}_{ck}
&=
\left[
\begin{array}{c|c|c}
\bar{\mathbf{v}}_{ck,1} &
\cdots &
\bar{\mathbf{v}}_{ck,M}
\end{array}
\right]_{M(K+1) \times M} \\
\bar{\mathbf{U}}_{ck}
&=
\left[
\begin{array}{c|c|c}
\bar{\mathbf{u}}_{ck,1} &
\cdots &
\bar{\mathbf{u}}_{ck,M}
\end{array}
\right]_{M(K+1) \times M}.
\end{align}

Now we proceed to show the achievable sum DoF of the scheme is $\frac{2KM}{K+1}$.
Since alignment of inter-cell interference is ensured by (\ref{eqn:v_design1}), the main task of the proof is to show that all signals are distinguishable at the intended receiver, despite the block-diagonal structure of the channel matrix given in (\ref{eqn:extended_H}).
The following
lemma establishes the linear independence required by the
proposed scheme.
\begin{lemma}
Let
\begin{align}
\mathbf{R}^{[1]}
&=
\left[
\begin{array}{c|c|c}
\bar{\mathbf{r}}^{[1]}_1 &
\cdots &
\bar{\mathbf{r}}^{[1]}_M
\end{array}
\right] \\
\mathbf{S}^{[1]}
& =
\left[
\begin{array}{c|c|c}
\bar{\mathbf{H}}_{11}^{[1]}
\bar{\mathbf{V}}_{11} &
\cdots &
\bar{\mathbf{H}}_{1K}^{[1]}
\bar{\mathbf{V}}_{1K}
\end{array}
\right].
\end{align}
Then the matrix
$\left[\mathbf{R}^{[1]} ~ | ~ \mathbf{S}^{[1]} \right] \in \mathbb{C}^{M(K+1) \times M(K+1)}$ is full rank with probability one.
\end{lemma}
\qquad \emph{Proof:}
We first show that the $MK$ column vectors of $\mathbf{S}^{[1]}$ are linearly independent with probability one.
Consider the vector equation
\begin{align}
\label{eqn:independent_step1}
\sum_{m=1}^M
\sum_{k=1}^K
c_{k,m}
\bar{\mathbf{H}}_{1k}^{[1]}
\bar{\mathbf{v}}_{1k,m}
=
\mathbf{0}_{M(K+1)}
\end{align}
where $c_{k,m}$ are scalars.
We aim to show that all $MK$ scalars $c_{k,m}$ in (\ref{eqn:independent_step1}) are zero.
Using the property of block-diagonal matrices, we can write $\bar{\mathbf{H}}_{1k}^{[1]} \bar{\mathbf{v}}_{1k,m}$ as
\begin{align}
\label{eqn:independent_step2}
\bar{\mathbf{H}}_{1k}^{[1]}
\bar{\mathbf{v}}_{1k,m}
=
\frac{\sqrt{(K+1)P}}{\sqrt{M} \big|\big| \bar{\mathbf{H}}_{1k}^{[2]^{-1}}
\bar{\mathbf{r}}^{[2]}_m \big|\big|}
\left[
\begin{array}{c}
\mathbf{H}_{1k}^{[1]}
\mathbf{H}_{1k}^{{[2]}^{-1}}
\mathbf{r}^{[2]}_m(1) \\
\vdots \\
\mathbf{H}_{1k}^{[1]}
\mathbf{H}_{1k}^{{[2]}^{-1}}
\mathbf{r}^{[2]}_m(T)
\end{array}
\right]
\end{align}
where $\mathbf{r}^{[2]}_m(t)$ is the segment of $\bar{\mathbf{r}}^{[2]}_m$ at time $t$ as the similar notation given in (\ref{eqn:scheme_IMAC}) for $t=1,\ldots,T$.
Substituting (\ref{eqn:independent_step2}) into (\ref{eqn:independent_step1}), we obtain the equivalent condition of (\ref{eqn:independent_step1}) for each time slot $t$ as
\begin{align}
\label{eqn:independent_step3}
\sum_{m=1}^M
\sum_{k=1}^K c_{k,m}
\mathbf{H}_{1k}^{[1]}
\mathbf{H}_{1k}^{{[2]}^{-1}}
\mathbf{r}_m^{[2]}(t)
=
\mathbf{0}_M
\end{align}
for $t=1,\dots,K+1$.
Note that the normalization terms in (\ref{eqn:v_design1}) and (\ref{eqn:independent_step2}) are ignored in (\ref{eqn:independent_step3}) for simple exploration without effecting the result.
Defining $\mathbf{F}_k$ as
\begin{equation}
\label{eqn:F}
\mathbf{F}_k
\triangleq
\mathbf{H}_{1k}^{[1]}
\mathbf{H}_{1k}^{[2]^{-1}}
\end{equation}
and rearranging (\ref{eqn:independent_step3}) in matrix form, we have
\begin{align}
\label{eqn:independent_step4}
\left[
\begin{array}{c|c|c}
\sum_{k=1}^{K} c_{k,1}
\mathbf{F}_k &
\cdots &
\sum_{k=1}^{K} c_{k,M}
\mathbf{F}_k
\end{array}
\right]
\mathbf{Q}
=
\mathbf{O}_{M\times(K+1)}
\end{align}
where $\mathbf{Q} \in \mathbb{C}^{M^2 \times (K+1)}$ is given as
\begin{align}
\label{eqn:independent_step5}
\mathbf{Q}
=
\left[
\begin{array}{c|c|c}
\mathbf{r}^{[2]}_1(1) &
\cdots &
\mathbf{r}^{[2]}_1(K+1)
\\
\vdots &
\ddots &
\vdots
\\
\mathbf{r}^{[2]}_M(1) &
\cdots &
\mathbf{r}^{[2]}_M(K+1)
\end{array}
\right].
\end{align}

Let's first consider the cases of $K=M^2-1$ and $M^2$. Since $\mathbf{Q}$ is a generic matrix by design, it is invertible and right invertible with probability one for $K=M^2-1$ and $M^2$, respectively.
Multiplying both sides of (\ref{eqn:independent_step5}) from the right by the inverse or the right inverse of $\mathbf{Q}$ gives us
\begin{align}
\label{eqn:independent_step6}
\left[
\begin{array}{c|c|c}
\sum_{k=1}^{K} c_{k,1}
\mathbf{F}_k &
\cdots &
\sum_{k=1}^{K} c_{k,M}
\mathbf{F}_k
\end{array}
\right]
=
\mathbf{O}_{M \times M^2},
\end{align}
which implies
\begin{align}
\label{eqn:independent_step7}
\sum_{k=1}^{K} c_{k,m}
\mathbf{F}_k
= \mathbf{O}_{M \times M}
\end{align}
for $m=1,\dots,M$. By the fact that the set of all $M \times M$ matrices can be considered as an $M^2$-dimensional vector space, and by the assumption that all channel matrices are generic, the $K$ matrices
$\mathbf{F}_{k}, ~k=1,\dots,K$, where $K \leq M^2$,
are linearly independent with probability one. Thus, (\ref{eqn:independent_step7}) implies that $c_{1,1} = \cdots = c_{K,M} = 0$.
Thus, all $MK$ column vectors of $\mathbf{S}^{[1]}$ are linearly independent with probability one.

Consider the remaining case that $K=2,\dots,M^2-2$, where $\mathbf{Q}$ is not invertible.
Let
$\{ \mathbf{q}^{\perp}_1, \ldots, \mathbf{q}^{\perp}_{M^2-(K+1)} \}$
be a basis of the orthogonal space of the column space of $\mathbf{Q}$, implying
$\mathbf{q}^{\perp}_i \mathbf{Q} = \mathbf{O}_{1 \times (K+1)}.$
Then (\ref{eqn:independent_step4}) implies that there exist $\alpha_{i,j} \in \mathbb{C}$ such that
\begin{align}
\label{eqn:independent_step8_and_half}
\left[
\begin{array}{c}
\sum_{i=1}^{M^2-(K+1)} \alpha_{i,1} \mathbf{q}^{\perp}_i \\
\vdots \\
\sum_{i=1}^{M^2-(K+1)} \alpha_{i,M} \mathbf{q}^{\perp}_i
\end{array}
\right]
=
\mathbf{Q}^{\perp}
\end{align}
where
\begin{align}
\label{eqn:independent_step9}
\mathbf{Q}^{\perp}
=
\left[
\begin{array}{c|c|c}
\sum_{k=1}^{K} c_{k,1}
\mathbf{F}_k &
\cdots &
\sum_{k=1}^{K} c_{k,M}
\mathbf{F}_k
\end{array}
\right]
\end{align}
and $\mathbf{Q}^{\perp}$ satisfies $\mathbf{Q}^{\perp} \mathbf{Q} = \mathbf{O}_{M \times (K+1)}$.
Applying the $\textrm{vec}(\cdot)$ operation to both sides of (\ref{eqn:independent_step8_and_half}) and with some manipulations, we have (\ref{eqn:independent_step10}), which is given at the bottom of this page.
\begin{figure*}[b]
\hrule
\vspace*{5pt}
\begin{align}
\label{eqn:independent_step10}
\mathbf{0}_{M^3}
& =
\sum_{k=1}^{K} c_{k,1}
\textrm{vec}
\left(
\left[
\begin{array}{c|c|c|c}
\mathbf{F}_{k} & \mathbf{O}_{M \times M} & \cdots & \mathbf{O}_{M \times M}
\end{array}
\right]
\right)
+
\cdots
+
\sum_{k=1}^{K} c_{k,M}
\textrm{vec}
\left(
\left[
\begin{array}{c|c|c|c}
\mathbf{O}_{M \times M} & \cdots & \mathbf{O}_{M \times M} & \mathbf{F}_{k}
\end{array}
\right]
\right) \nonumber \\
& \quad -
\sum_{i=1}^{M^2-(K+1)}
\alpha_{i,1}
\textrm{vec}
\left(
\left[
\begin{array}{c}
\mathbf{q}^{\perp}_i \\
\mathbf{O}_{1 \times M^2} \\
\vdots \\
\mathbf{O}_{1 \times M^2}
\end{array}
\right]
\right)
-
\cdots
-
\sum_{i=1}^{M^2-(K+1)}
\alpha_{i,M}
\textrm{vec}
\left(
\left[
\begin{array}{c}
\mathbf{O}_{1 \times M^2} \\
\vdots \\
\mathbf{O}_{1 \times M^2} \\
\mathbf{q}^{\perp}_i
\end{array}
\right]
\right).
\end{align}
\end{figure*}
By the fact that $\mathbf{F}_1, \ldots, \mathbf{F}_K$, and $\mathbf{Q}$ are outcomes of (statistically) independent random matrices, the fact that $\mathbf{q}^{\perp}_1, \ldots, \mathbf{q}^{\perp}_{M^2-(K+1)}$ are linearly independent, and the fact that there are total $M^3 - M$ vectors in the $M^3$-dimensional vector space, we obtain that $c_{1,1} = \cdots = c_{K,M} = \alpha_{1,1} = \cdots = \alpha_{M^2-(K+1),M} = 0$. Thus, the rank of $\mathbf{S}^{[1]}$ is $MK$ with probability one.
Since the $M$ column vectors of $\mathbf{R}^{[1]}$ are generic vectors in $\mathbb{C}^{M(K+1)}$, along with the fact that $\mathbf{S}^{[1]}$ and $\mathbf{R}^{[1]}$ are outcomes of two (statistically) independent random matrices, we have $\mathrm{rank}(\left[\mathbf{R}^{[1]} ~ | ~ \mathbf{S}^{[1]} \right]) = M(K+1)$.
\hfill$\blacksquare$

\vspace{2pt}

The following remark discusses the key ideas in the proof.
\begin{remark}
The relation between the size of the channel matrix and the maximum number of the active users can be observed in (\ref{eqn:independent_step7}). Note that here the idea of channel diversity is related directly to the channel matrix and is utilized to provide the achievability proof, while the extended channel matrix and converse proof are considered in \cite{Li_Jafarkhani_Jafar}.
\end{remark}

\subsection{Proof of Converse}
\label{sec:proof_ccs_up}
The following lemma applies to all possible precoding matrices operated in the extended signal space $\mathbb{C}^{MT}$ with an arbitrary number of time slots $T$.
\begin{lemma}
\label{lmm:R}
For all precoding matrices of arbitrary sizes $\bar{\mathbf{V}}_{1k}$, $k=1,\dots,K$, in cell 1, if
\begin{align}
\label{eqn:lmm_R}
\mathrm{rank}
\left(
\left[
\begin{array}{c|c|c}
\bar{\mathbf{H}}_{11}^{[2]}
\bar{\mathbf{V}}_{11} &
\cdots &
\bar{\mathbf{H}}_{1K}^{[2]}
\bar{\mathbf{V}}_{1K}
\end{array}
\right]
\right) = R,
\end{align}
then with probability one
\begin{align}
\label{eqn:lmm_M2R}
\mathrm{rank}
\left(
\left[
\begin{array}{c|c|c}
\bar{\mathbf{H}}_{11}^{[1]}
\bar{\mathbf{V}}_{11} &
\cdots &
\bar{\mathbf{H}}_{1K}^{[1]}
\bar{\mathbf{V}}_{1K}
\end{array}
\right]
\right) \leq M^2R.
\end{align}
\end{lemma}
\qquad \emph{Proof:} Equation (\ref{eqn:lmm_R}) implies that there exist $a_{1k,mr} \in \mathbb{C}$ and $\bar{\mathbf{r}}_1^{[2]},\ldots,\bar{\mathbf{r}}_R^{[2]} \in \mathbb{C}^{MT}$ such that the $m^{th}$ precoding vector of user $k$ in cell one, i.e., the $m^{th}$ column of $\bar{\mathbf{V}}_{1k}$, satisfies
\begin{equation}
\label{eqn:interference}
\bar{\mathbf{H}}_{1k}^{[2]} \bar{\mathbf{v}}_{1k,m}
=
\sum_{r=1}^{R} a_{1k,mr} \bar{\mathbf{r}}_r^{[2]}.
\end{equation}
Applying (\ref{eqn:interference}) to all $n_{1k}$ column vectors of $\bar{\mathbf{H}}_{1k}^{[2]} \bar{\mathbf{V}}_{1k}$, we could write $\bar{\mathbf{H}}_{1k}^{[2]} \bar{\mathbf{V}}_{1k}$ as
\begin{align}
\bar{\mathbf{H}}_{1k}^{[2]} \bar{\mathbf{V}}_{1k}
& =
\left[
\begin{array}{c|c|c}
\sum_{r=1}^{R} a_{1k,1r} \bar{\mathbf{r}}_r^{[2]} &
\cdots &
\sum_{r=1}^{R} a_{1k,n_{1k}r} \bar{\mathbf{r}}_r^{[2]}
\end{array}
\right] \nonumber \\
& =
\sum_{r=1}^{R}
\left[
\begin{array}{c|c|c}
a_{1k,1r} \bar{\mathbf{r}}_r^{[2]} &
\cdots &
a_{1k,n_{1k}r} \bar{\mathbf{r}}_r^{[2]}
\end{array}
\right].
\end{align}
Since $\bar{\mathbf{H}}_{1k}^{[2]}$ is invertible with probability one, we can use the property of block-diagonal matrices to obtain
\begin{align}
\bar{\mathbf{H}}_{1k}^{[1]} \bar{\mathbf{V}}_{1k}
& =
\sum_{r=1}^{R}
\bar{\mathbf{H}}_{1k}^{[1]}
\bar{\mathbf{H}}_{1k}^{{[2]}^{-1}}
\left[
\begin{array}{c|c|c}
a_{1k,1r} \bar{\mathbf{r}}_r^{[2]} &
\cdots &
a_{1k,n_{1k}r} \bar{\mathbf{r}}_r^{[2]}
\end{array}
\right] \nonumber \\
& =
\sum_{r=1}^{R}
\left[
\begin{array}{ccc}
a_{1k,1r}
\mathbf{F}_k
\mathbf{r}_r^{[2]}(1) &
\cdots &
a_{1k,n_{1k}r}
\mathbf{F}_k
\mathbf{r}_r^{[2]}(1) \\
a_{1k,1r}
\mathbf{F}_k
\mathbf{r}_r^{[2]}(2) &
\cdots &
a_{1k,n_{1k}r}
\mathbf{F}_k
\mathbf{r}_r^{[2]}(2) \\
\vdots &
\ddots &
\vdots \\
a_{1k,1r}
\mathbf{F}_k
\mathbf{r}_r^{[2]}(T) &
\cdots &
a_{1k,n_{1k}r}
\mathbf{F}_k
\mathbf{r}_r^{[2]}(T)
\end{array}
\right]
\label{eqn:HV}
\end{align}
where $\mathbf{F}_k$ is given in (\ref{eqn:F}) and $\mathbf{r}^{[2]}_r(t)$ is the segment of $\bar{\mathbf{r}}^{[2]}_r$ at time $t$ as the similar notation given in (\ref{eqn:scheme_IMAC}) for $t=1,\ldots,T$.
Using (\ref{eqn:HV}), we can write
\begin{equation}
\mathbf{S}^{[1]}
\triangleq
\left[
\begin{array}{c|c|c}
\bar{\mathbf{H}}_{11}^{[1]}
\bar{\mathbf{V}}_{11} &
\cdots &
\bar{\mathbf{H}}_{1K}^{[1]}
\bar{\mathbf{V}}_{1K}
\end{array}
\right]
=\sum_{r=1}^{R} \mathbf{A}_r
\end{equation}
where $\mathbf{A}_r$ is the $MT \times \sum_{k=1}^{K} n_{1k}$ matrix given at the bottom of this page.

\begin{figure*}[b]
\hrule
\vspace*{5pt}
\begin{align}
\label{eqn:Ar}
\mathbf{A}_r
&=
\left[
\begin{array}{ccc|c|ccc}
a_{11,1r}
\mathbf{F}_1
\mathbf{r}_r^{[2]}(1) &
\cdots &
a_{11,n_{11}r}
\mathbf{F}_1
\mathbf{r}_r^{[2]}(1) &
\cdots &
a_{1K,1r}
\mathbf{F}_K
\mathbf{r}_r^{[2]}(1) &
\cdots &
a_{1K,n_{1K}r}
\mathbf{F}_K
\mathbf{r}_r^{[2]}(1) \\
a_{11,1r}
\mathbf{F}_1
\mathbf{r}_r^{[2]}(2) &
\cdots &
a_{11,n_{11}r}
\mathbf{F}_1
\mathbf{r}_r^{[2]}(2) &
\cdots &
a_{1K,1r}
\mathbf{F}_K
\mathbf{r}_r^{[2]}(2) &
\cdots &
a_{1K,n_{1K}r}
\mathbf{F}_K
\mathbf{r}_r^{[2]}(2) \\
\vdots &
\ddots &
\vdots &
\cdots &
\vdots &
\ddots &
\vdots \\
a_{11,1r}
\mathbf{F}_1
\mathbf{r}_r^{[2]}(T) &
\cdots &
a_{11,n_{11}r}
\mathbf{F}_1
\mathbf{r}_r^{[2]}(T) &
\cdots &
a_{1K,1r}
\mathbf{F}_K
\mathbf{r}_r^{[2]}(T) &
\cdots &
a_{1K,n_{1K}r}
\mathbf{F}_K
\mathbf{r}_r^{[2]}(T)
\end{array}
\right]
\end{align}
\vspace*{5pt}
\hrule
\vspace*{5pt}
\begin{align}
\label{eqn:Brm}
\mathbf{B}_{rm}
&=
\left[
\begin{array}{ccc|c|ccc}
b_{1rm}
a_{11,1r}
\mathbf{F}_1
\mathbf{e}_m &
\cdots &
b_{1rm}
a_{11,n_{11}r}
\mathbf{F}_1
\mathbf{e}_m &
\cdots &
b_{1rm}
a_{1K,1r}
\mathbf{F}_K
\mathbf{e}_m &
\cdots &
b_{1rm}
a_{1K,n_{1K}r}
\mathbf{F}_K
\mathbf{e}_m \\
b_{2rm}
a_{11,1r}
\mathbf{F}_1
\mathbf{e}_m &
\cdots &
b_{2rm}
a_{11,n_{11}r}
\mathbf{F}_1
\mathbf{e}_m &
\cdots &
b_{2rm}
a_{1K,1r}
\mathbf{F}_K
\mathbf{e}_m &
\cdots &
b_{2rm}
a_{1K,n_{1K}r}
\mathbf{F}_K
\mathbf{e}_m \\
\vdots &
\ddots &
\vdots &
\cdots &
\vdots &
\ddots &
\vdots \\
b_{Trm}
a_{11,1r}
\mathbf{F}_1
\mathbf{e}_m &
\cdots &
b_{Trm}
a_{11,n_{11}r}
\mathbf{F}_1
\mathbf{e}_m &
\cdots &
b_{Trm}
a_{1K,1r}
\mathbf{F}_K
\mathbf{e}_m &
\cdots &
b_{Trm}
a_{1K,n_{1K}r}
\mathbf{F}_K
\mathbf{e}_m
\end{array}
\right]
\end{align}
\end{figure*}

Let $b_{trm}$ denote the $m^{th}$ element of $\mathbf{r}_r^{[2]}(t)$. Then $\mathbf{F}_k \mathbf{r}_r^{[2]}(t)$ can be expressed as
\begin{align}
\mathbf{F}_k
\mathbf{r}_r^{[2]}(t)
=
\mathbf{F}_k
\left[
~ b_{tr1} ~\cdots ~ b_{trM} ~
\right]^t
=
\label{eqn:rr}
\sum_{m=1}^M
b_{trm}
\mathbf{F}_k
\mathbf{e}_m
\end{align}
where $\mathbf{e}_m \in \mathbb{C}^M$ is the elementary vector whose elements are all zero except that the $m^{th}$ element is 1.
Substituting (\ref{eqn:rr}) into (\ref{eqn:Ar}) and reorganizing the summation, we have
\begin{equation}
\mathbf{A}_r = \sum_{m=1}^M \mathbf{B}_{rm}
\end{equation}
where $\mathbf{B}_{rm}$ is the $MT \times \sum_{k=1}^{K} n_{1k}$ matrix given at the bottom of the previous page.

Note that the first $M$ rows of $\mathbf{B}_{rm}$ are proportional to the next $M$ rows, i.e., row $M+1$ to row $2M$, with ratio $\frac{b_{1rm}}{b_{2rm}}$. Similar observations can be made for all the following rows. Thus, we have
\begin{equation}
\textrm{rank}
\left(
\mathbf{B}_{rm}
\right) \leq M.
\end{equation}
Finally, $\textrm{rank}(\mathbf{S}^{[1]})$ is upperbounded as follows.
\begin{equation}
\textrm{rank}
(
\mathbf{S}^{[1]}
)
\stackrel{(a)}{\leq}
\sum_{r=1}^R
\sum_{m=1}^M
\textrm{rank}
\left(
\mathbf{B}_{rm}
\right)
\leq
M^2 R
\end{equation}
where $(a)$ follows from the fact that $\textrm{rank}(\mathbf{A}+\mathbf{B}) \leq \textrm{rank}(\mathbf{A}) + \textrm{rank}(\mathbf{B})$. This concludes the proof of Lemma \ref{lmm:R}.
\hfill$\blacksquare$

\vspace{2pt}

\begin{figure}[t]
\begin{center}
\includegraphics[width=240pt, trim=10 15 11 0, clip]{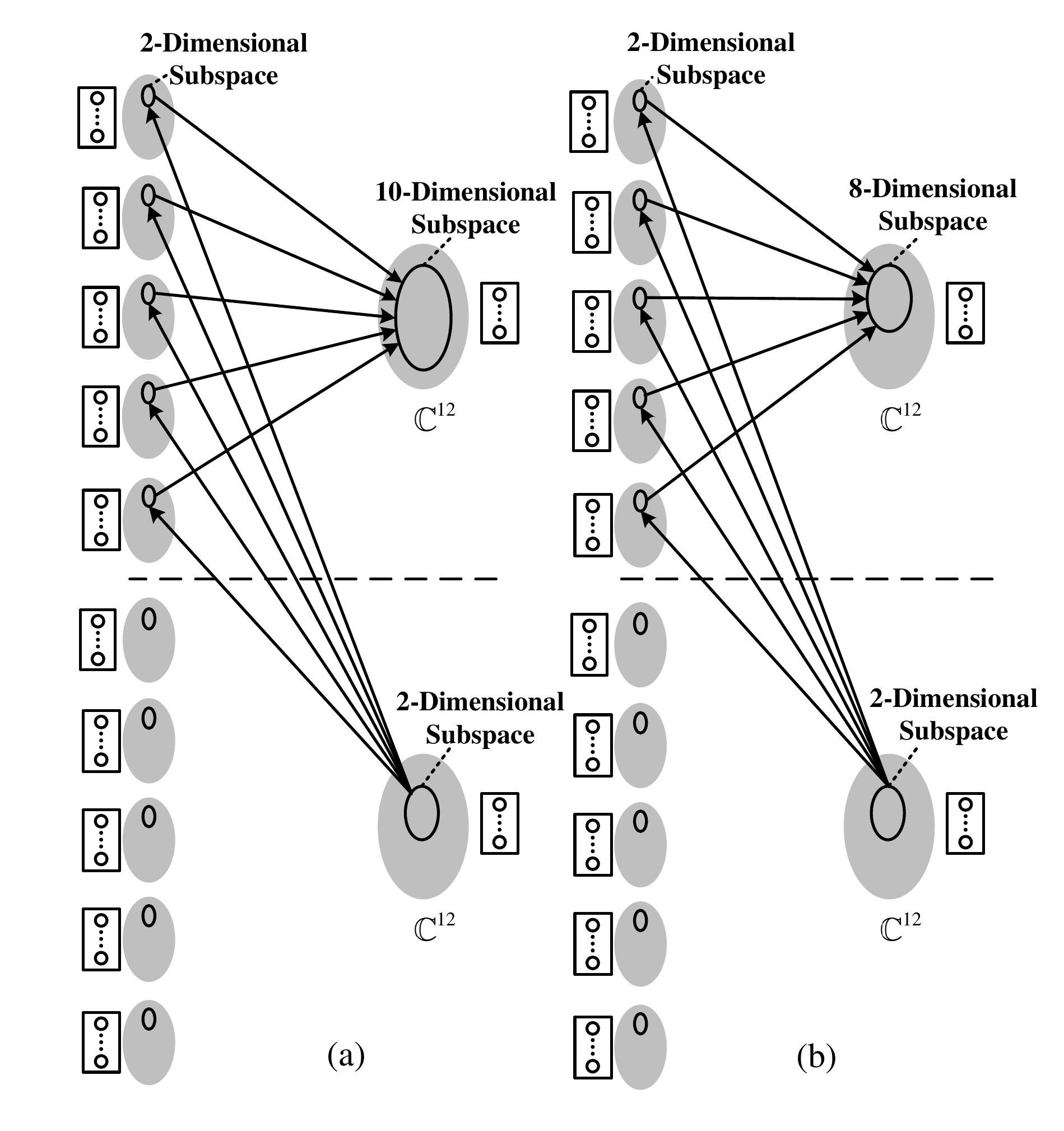}
\caption{Illustration of Lemma \ref{lmm:R} that shows the relation between the interference dimensions observed by receiver 2 and the signal dimensions observed by receiver 1.  Part (a) shows the signal space for $M=12$. Part (b) shows the extended signal space for $M=2$ with 6-symbol extension.}\label{fig:up}
\end{center}
\end{figure}

Lemma \ref{lmm:R} is illustrated in Fig. \ref{fig:up}. Note that Lemma \ref{lmm:R} indicates that for all precoding vectors in cell $1$, the ratio of signal dimensions observed at receiver $1$ to the interference dimensions observed at receiver $2$ is upperbounded by $RM^2 / R = M^2$. Thus, simple calculation shows that equal interference dimensions at both receivers lead to the maximum sum DoF achieved by linear IA, which is upperbounded as
\begin{equation}
d_{\textrm{f},\textrm{css}}(K,M)
\leq
\frac{2 \cdot \frac{RM^2}{RM^2+R} \cdot TM}{T}
=
\frac{2M^3}{M^2+1}.
\end{equation}
This concludes the proof of Theorem \ref{thm:d_f_up}.
\begin{remark}
Unlike the DoF upper bounds in \cite{Li_Jafarkhani_Jafar}, there is no assumption on the intersection of interference subspaces and the number of data streams sent by each transmitter in Lemma \ref{lmm:R} and Theorem \ref{thm:d_f_up}.
\end{remark}

\section{Proof for ACS Systems}
\label{sec:proof_acs}
In this section, we prove the achievability and converse of Theorem \ref{thm:d_f_acs}. We begin with the achievability.

\subsection{Proof of Achievability}
\label{sec:proof_acs_low}
The achievability proof of Theorem \ref{thm:d_f_acs} is similar to that of Theorem \ref{thm:d_f_ccs}.
Thus, we only describe several key steps to point out the differences between the CSS scheme and the ACS scheme.
Let $T = K+1$, and choose the $4M$ reference vectors
$\widetilde{\mathbf{r}}^{[1]}_1, \ldots, \widetilde{\mathbf{r}}^{[1]}_{2M}, \widetilde{\mathbf{r}}^{[2]}_1, \ldots, \widetilde{\mathbf{r}}^{[2]}_{2M} \in \mathbb{R}^{2M(K+1)}$
by letting the elements of the vectors be outcomes of independent continuous random variables.
Let the $m^{th}$ precoding vector of user $k$ in cell $c$ be
\begin{align}
\label{eqn:v_design1_acs}
\widetilde{\mathbf{v}}_{ck,m} &=
\frac{\sqrt{(K+1)P}}{\sqrt{2M} \big|\big| \widetilde{\mathbf{H}}_{ck}^{[\bar{c}]^{-1}}
\widetilde{\mathbf{r}}^{[\bar{c}]}_m \big|\big|}
\widetilde{\mathbf{H}}_{ck}^{[\bar{c}]^{-1}}
\widetilde{\mathbf{r}}^{[\bar{c}]}_m.
\end{align}
for $m=1,\ldots,2M$ and $(c, \bar{c}) = (1, 2), (2, 1)$.
The next key step is to show that the union of the set reference vectors
$\widetilde{\mathcal{R}}_1 = \{ \widetilde{\mathbf{r}}^{[1]}_1, \ldots, \widetilde{\mathbf{r}}^{[1]}_{2M} \}$
and the set of intended signals
$\left\{
\widetilde{\mathbf{H}}^{[1]}_{1k} \widetilde{\mathbf{v}}_{1k,m}:
(m,k) = (1, 1), \ldots, (2M,K)
\right\}$
is a linearly independent set.
The main task of this step is to show that the vector equation
\begin{align}
\label{eqn:independent_step1_acs}
\sum_{m=1}^{2M}
\sum_{k=1}^K
\widetilde{c}_{k,m}
\widetilde{\mathbf{H}}_{1k}^{[1]}
\widetilde{\mathbf{v}}_{1k,m}
=
\mathbf{0}_{2M(K+1)}
\end{align}
implies that all $2MK$ scalars $\widetilde{c}_{k,m}$ are zero.
The remaining part of the proof follows the similar steps in Section \ref{sec:proof_ccs_low} and is omitted here to avoid repetition.

\subsection{Proof of Converse}
\label{sec:proof_acs_up}
We prove the converse of Theorem \ref{thm:d_f_acs} in this section.
Although the proof follows some similar steps given in Section \ref{sec:proof_ccs_up}, there are also several new important ingredients, which are the keys to the proof. The converse of Theorem \ref{thm:d_f_acs} is stated in the following theorem.
\begin{theorem}
\label{thm:d_f_up_acs}
\emph{(Converse:)}
The feasible sum DoF satisfies
\begin{align}
\label{eqn:d_f_up_acs}
d_{\textrm{f},\textrm{acs}}(K,M) \leq \frac{4M^3}{2M^2+1}.
\end{align}
\end{theorem}

To prove Theorem \ref{thm:d_f_up_acs}, we need the following lemma that applies to all precoding matrices operated in the ACS extend signal space $\mathbb{R}^{2MT}$ with an arbitrary number of time slots $T$.

\begin{lemma}
\label{lmm:R_ACS}
For all precoding matrices of arbitrary sizes $\widetilde{\mathbf{V}}_{1k}$, $k=1,\dots,K$, in cell 1, if
\begin{align}
\label{eqn:lmm_R_ACS}
\mathrm{rank}
\left(
\left[
\begin{array}{c|c|c}
\widetilde{\mathbf{H}}_{11}^{[2]}
\widetilde{\mathbf{V}}_{11} &
\cdots &
\widetilde{\mathbf{H}}_{1K}^{[2]}
\widetilde{\mathbf{V}}_{1K}
\end{array}
\right]
\right) = R,
\end{align}
then with probability one
\begin{align}
\label{eqn:lmm_2M2R}
\mathrm{rank}
\left(
\left[
\begin{array}{c|c|c}
\widetilde{\mathbf{H}}_{11}^{[1]}
\widetilde{\mathbf{V}}_{11} &
\cdots &
\widetilde{\mathbf{H}}_{11}^{[1]}
\widetilde{\mathbf{V}}_{11}
\end{array}
\right]
\right) \leq 2M^2R.
\end{align}
\end{lemma}
\qquad \emph{Proof:}
Equation (\ref{eqn:lmm_R_ACS}) implies that there exist $\widetilde{a}_{1k,mr} \in \mathbb{R}$ and $\widetilde{\mathbf{r}}_{1}^{[2]},\ldots,\widetilde{\mathbf{r}}_{R}^{[2]} \in \mathbb{R}^{2MT}$ such that the $m^{th}$ precoding vector of user $k$ in cell one, i.e., the $m^{th}$ column of $\widetilde{\mathbf{V}}_{1k}$, satisfies
\begin{equation}
\label{eqn:interference_ACS}
\widetilde{\mathbf{H}}_{1k}^{[2]} \widetilde{\mathbf{v}}_{1k,m}
=
\sum_{r=1}^{R} \widetilde{a}_{1k,mr} \widetilde{\mathbf{r}}_{r}^{[2]}.
\end{equation}
Applying (\ref{eqn:interference_ACS}) to all $n_{1k}$ column vectors of $\widetilde{\mathbf{H}}_{1k}^{[2]} \widetilde{\mathbf{V}}_{1k}$ and by the fact that $\widetilde{\mathbf{H}}_{1k}^{[2]}$ is block-diagonal and invertible with probability one, we obtain
\begin{equation}
\label{eqn:S_tilde}
\widetilde{\mathbf{S}}^{[1]}
\triangleq
\left[
\begin{array}{c|c|c}
\widetilde{\mathbf{H}}_{11}^{[1]}
\widetilde{\mathbf{V}}_{11} &
\cdots &
\widetilde{\mathbf{H}}_{1K}^{[1]}
\widetilde{\mathbf{V}}_{1K}
\end{array}
\right]
=
\sum_{r=1}^{R} \widetilde{\mathbf{A}}_r
\end{equation}
where $\widetilde{\mathbf{A}}_r$ is the $2MT \times \sum_{k=1}^{K} n_{1k}$ matrix of the form given in (\ref{eqn:Ar}) with each block $a_{1k,mr} \mathbf{F}_k \mathbf{r}_{r}^{[2]}(t)$ being replaced by $\widetilde{a}_{1k,mr} \check{\mathbf{F}}_k \check{\mathbf{r}}_{r}^{[2]}(t)$, where $\check{\mathbf{r}}_{r}^{[2]}(t)$ is the segment of $\widetilde{\mathbf{r}}_{r}^{[2]}$ at time $t$ as the similar notation given in (\ref{eqn:tilde_x}) and $\check{\mathbf{F}}_k$ is defined as
\begin{align}
\label{eqn:F_ACS}
\check{\mathbf{F}}_k \triangleq \check{\mathbf{H}}^{[1]}_{1k} \check{\mathbf{H}}^{{[2]}^{-1}}_{1k}.
\end{align}
Let $\widetilde{b}_{trm}$ denote the $m^{th}$ element of $\check{\mathbf{r}}_{r}^{[2]}(t)$. Then $\check{\mathbf{F}}_k \check{\mathbf{r}}_{r}^{[2]}(t)$ can be expressed as
\begin{align}
\label{eqn:rr_ACS}
\check{\mathbf{F}}_k
\check{\mathbf{r}}_{r}^{[2]}(t)
=
\sum_{m=1}^{2M}
\widetilde{b}_{trm}
\check{\mathbf{F}}_k
\mathbf{e}_m
\end{align}
where $\mathbf{e}_m \in \mathbb{R}^{2M}$ is the elementary vector whose elements are all zero except that the $m^{th}$ element is 1.
Now we can express $\widetilde{\mathbf{A}}_r$ as
\begin{align}
\label{eqn:Ar_tilde}
\widetilde{\mathbf{A}}_r = \sum_{m=1}^{2M} \widetilde{\mathbf{B}}_{rm}
\end{align}
where $\widetilde{\mathbf{B}}_{rm}$ is the $2MT \times \sum_{k=1}^{K} n_{1k}$ matrix of the form given in (\ref{eqn:Brm}) with each block
$b_{trm} a_{1k,mr} \mathbf{F}_k \mathbf{e}_m$
being replaced by
$\widetilde{b}_{trm} \widetilde{a}_{1k,mr} \check{\mathbf{F}}_k \mathbf{e}_m$.
Note that unlike the CSS case, where $\textrm{rank}(\mathbf{B}_{rm})$ is upperbounded by $M$, here we have
\begin{align}
\label{eqn:Brm_rank_up}
\textrm{rank}
\left(
\widetilde{\mathbf{B}}_{rm}
\right) \leq 2M.
\end{align}

From (\ref{eqn:F_ACS}), we can see that $\check{\mathbf{F}}_k$ and $\check{\mathbf{H}}^{[r]}_{ck}$ have the same structure shown in (\ref{eqn:extended_H_acs_structure}).
An important property of this structure is that the first element in the second column is the negate of the second element in the first column, and the second element in the second column is the same as the first element in the first column, i.e.,
\begin{align}
(\check{\mathbf{F}}_k \mathbf{e}_2)_1
=
- (\check{\mathbf{F}}_k \mathbf{e}_1)_2,
~~
(\check{\mathbf{F}}_k \mathbf{e}_2)_2
=
(\check{\mathbf{F}}_k \mathbf{e}_1)_1.
\end{align}
Generalizing this observation, we have
\begin{align}
\label{eqn:Fe}
\check{\mathbf{F}}_k \mathbf{e}_{m+1}
=
\mathbf{P}_M
\check{\mathbf{F}}_k \mathbf{e}_m
\end{align}
for $m \in \{1, 3, \dots, 2M-1 \}$ and where
\begin{align}
\mathbf{P}_M
\triangleq
\textrm{blck}
\left(
\left[
\begin{array}{cc}
0 & -1 \\
1 & 0
\end{array}
\right],
\ldots,
\left[
\begin{array}{cc}
0 & -1 \\
1 & 0
\end{array}
\right]
\right)
\in \mathbb{R}^{2M \times 2M}.
\end{align}
Using (\ref{eqn:Fe}) to relate columns of
$\widetilde{\mathbf{B}}_{r(m+1)}$
with those of
$\widetilde{\mathbf{B}}_{rm}$,
we obtain
\begin{align}
\label{eqn:Brm_plus_one}
\widetilde{\mathbf{B}}_{r(m+1)} = \widetilde{\mathbf{P}}_{(m+1)mr}
\widetilde{\mathbf{B}}_{rm}
\end{align}
for $m \in \{1, 3, \dots, 2M-1 \}$ and where the $2MT \times 2MT$ matrix
$\widetilde{\mathbf{P}}_{(m+1)mr}$
is defined as
\begin{align}
\widetilde{\mathbf{P}}_{(m+1)mr}
\triangleq
\textrm{blck}
\left(
\frac{\widetilde{b}_{1r(m+1)}}{\widetilde{b}_{1rm}}
\mathbf{P}_M,
\ldots,
\frac{\widetilde{b}_{Tr(m+1)}}{\widetilde{b}_{Trm}}
\mathbf{P}_M
\right).
\end{align}
Based on (\ref{eqn:Ar_tilde}) and (\ref{eqn:Brm_plus_one}), we can rewrite $\widetilde{\mathbf{A}}$ as follows.
\begin{align}
\widetilde{\mathbf{A}}_r
&=
\sum_{m=1}^{2M}
\widetilde{\mathbf{B}}_{rm} \\
&=
\sum_{m=1}^{M}
\left(
\widetilde{\mathbf{B}}_{r(2m-1)}
+
\widetilde{\mathbf{B}}_{r(2m)}
\right) \\
&=
\label{eqn:Ar_tilde_rewrite}
\sum_{m=1}^{M}
\left(
\mathbf{I}_{2MT}
+
\widetilde{\mathbf{P}}_{(2m)(2m-1)r}
\right)
\widetilde{\mathbf{B}}_{r(2m-1)}.
\end{align}
Thus, substituting (\ref{eqn:Ar_tilde_rewrite}) into (\ref{eqn:S_tilde}), we can upperbound $\textrm{rank}(\widetilde{\mathbf{S}}^{[1]})$
as follows.
\begin{align}
\textrm{rank}
(\widetilde{\mathbf{S}}^{[1]})
& \stackrel{(a)}{\leq}
\sum_{r=1}^{R}
\sum_{m=1}^{M}
\textrm{rank}
\left(
\left(
\mathbf{I}
+
\widetilde{\mathbf{P}}_{(2m)(2m-1)r}
\right)
\widetilde{\mathbf{B}}_{r(2m-1)}
\right) \nonumber \\
& \stackrel{(b)}{\leq}
\sum_{r=1}^{R}
\sum_{m=1}^{M}
\textrm{rank}
\left(
\widetilde{\mathbf{B}}_{r(2m-1)}
\right) \nonumber \\
& \stackrel{(c)}{\leq}
2 M^2 R
\end{align}
where
$(a)$ follows from the fact that $\textrm{rank}(\mathbf{A}+\mathbf{B}) \leq \textrm{rank}(\mathbf{A}) + \textrm{rank}(\mathbf{B})$,
$(b)$ follows from the fact that $\textrm{rank}(\mathbf{AB}) \leq \textrm{rank}(\mathbf{B})$,
and $(c)$ follows from (\ref{eqn:Brm_rank_up}).
This concludes the proof of Lemma \ref{lmm:R_ACS}.
\hfill$\blacksquare$

\vspace{2pt}

Note that Lemma \ref{lmm:R_ACS} indicates that for all precoding vectors in cell $1$, the ratio of signal dimensions observed at receiver $1$ to the interference dimensions observed at receiver $2$ is upperbounded by $2M^2R / R = 2M^2$. Thus, simple calculation shows that equal interference dimensions at both receivers lead to the maximum sum DoF achieved by linear IA, which is upperbounded as
\begin{equation}
d_{\textrm{f},\textrm{acs}}(K,M)
\leq
\frac{2 \cdot \frac{2M^{2}R} {2M^2R+R} \cdot 2TM \cdot \frac{1}{2}}{T}=\frac{4M^3}{2M^2+1}
\end{equation}
This concludes the proof of Theorem \ref{thm:d_f_up_acs}.

\section{Extension to Parallel Channels}
\label{sec:parallel}
In this section, we extend our results to the parallel MIMO channel with $L$ subchannels, which is the model for the wideband systems using multi-carrier modulations such as orthogonal frequency division multiplexing (OFDM).
Consider the time-invariant two-cell MIMO IMAC with $L$ subchannels.  Assume that there are $K$ users in each cell and $M$ antennas at each node. The input-output equation of the channel is given as
\begin{align}
\label{eqn:parallel_ch}
\mathbf{y}^{[r]}_l(t)
=
\sum_{c=1}^2
\sum_{k=1}^K
\mathbf{H}_{ck,l}^{[r]}
\mathbf{x}_{ck,l}(t)
+
\mathbf{z}^{[r]}_l(t)
\end{align}
for $r = 1, 2$ and $l = 1, \ldots, L$,
where for the $l^{th}$ subchannel at the $t^{th}$ channel use,
$\mathbf{y}^{[r]}_l(t)$,
$\mathbf{z}^{[r]}_l(t)$ are the $M \times 1$ vectors
representing the channel output and additive white Gaussian noise at receiver $r$,
$\mathbf{H}_{ck,l}^{[r]}$ is
the $M \times M$ channel matrix from transmitter $k$ in cell $c$ to receiver $r$,
and $\mathbf{x}_{ck,l}(t)$ is the
$M \times 1$ channel input from transmitter $k$ in cell $c$, for
$r, c \in \{ 1, 2 \}$ and $k \in \{ 1, \ldots, K \}$.
The elements of $\mathbf{H}_{ck,l}^{[r]}$ are assumed to be outcomes of i.i.d. continuous random variables and do not change with $t$. The elements of $\mathbf{z}^{[r]}_l(t)$ are
i.i.d. (across space, time, and subchannels) circularly symmetric complex Gaussian random variables with zero mean and unit variance.
We assume that all channel matrices are known by all nodes in the channel.
The transmit power constraint is expressed as
\begin{align}
\label{eqn:power_cons_parallel}
\sum_{l=1}^L \mbox{E}[|| \mathbf{x}_{ck,l}(t) ||^2] \leq P.
\end{align}
The definitions of message set, capacity region, and sum DoF are the same as those given in Section \ref{sec:model}.

Note that (\ref{eqn:parallel_ch}) and (\ref{eqn:power_cons_parallel}) can be rewritten as the same input-output equation given in (\ref{eqn:channel}) and the same transmit power constraint given in (\ref{eqn:power_cons}) by jointly considering all $L$ subchannels and by letting
\begin{align}
\label{eqn:scheme_IMAC_parallel}
\mathbf{x}_{ck}(t) =
\left[
\begin{array}{c}
\mathbf{x}_{ck,1}(t) \\
\vdots \\
\mathbf{x}_{ck,L}(t)
\end{array}
\right], ~~
\mathbf{y}^{[r]}(t) =
\left[
\begin{array}{c}
\mathbf{y}^{[r]}_1(t) \\
\vdots \\
\mathbf{y}^{[r]}_L(t)
\end{array}
\right]
\end{align}
and
\begin{align}
\label{eqn:eq_H_parallel}
\mathbf{H}_{ck}^{[r]}
=
\textrm{blck}(\mathbf{H}_{ck,1}^{[r]}, \ldots, \mathbf{H}_{ck,L}^{[r]}).
\end{align}
Therefore, the idea of linear precoding/combining with symbol extension described in Section \ref{sec:model} and the proof methodologies described in Section \ref{sec:proof_ccs} and \ref{sec:proof_acs} can be applied to the equivalent MIMO channel, where there are $ML$ antennas in each node and where the channel matrix is block-diagonal with \emph{different} blocks, to obtain the following theorems.
\begin{theorem}
\label{thm:d_f_pl_ccs}
For the considered time-invariant two-cell parallel $K$-user IMAC with $L$ subchannels and $M$ antennas at each node, the largest achievable DoF provided by the linear IA scheme with symbol extensions is
\begin{align}
\label{eqn:d_f_pl_ccs}
d_{\textrm{f},\textrm{css}} =
\left\{
\begin{array}{ll}
2KML/(K+1) & \mathrm{if} ~ K \leq M^2L \\
2M^3L^2/(M^2L + 1) & \mathrm{if} ~ K > M^2L
\end{array}
\right.
\end{align}
for CSS systems, and
\begin{align}
\label{eqn:d_f_pl_acs}
d_{\textrm{f},\textrm{acs}} =
\left\{
\begin{array}{ll}
2KML/(K+1) & \mathrm{if} ~ K \leq 2M^2L \\
4M^3L^2/(2M^2L + 1) & \mathrm{if} ~ K > 2M^2L
\end{array}
\right.
\end{align}
for ACS systems.
\end{theorem}
\qquad \emph{Proof:}
The proof follows the similar steps in previous sections and is omitted here to avoid repetition.
\hfill $\blacksquare$

\section{Conclusion}
\label{sec:conclusion}
In this paper, we consider the time-invariant two-cell MIMO interfering multiple access channels and provide the exact characterization of the maximum achievable sum DoF under the constraint of using linear interference alignment scheme with symbol extensions. We show that, unlike the time-varying channels, the time-invariant channels impose a channel diversity constraint that arises from the block-diagonal structure of the symbol-extended channel matrices, where all blocks are the same. Our results explicitly indicate how this constraint restricts the maximum number of simultaneous active users in each cell, which in turn restricts the maximum linearly achievable sum DoF. To obtain these results, we propose a novel DoF upper bound, which applies to all possible precoding and combining matrices for arbitrary number of data streams and with arbitrary number of time slots. The proposed upper bound is based on a lemma that derives a rank ratio inequality, which is originally proposed for the time-varying $X$ channel \cite{Lashgari_Avestimehr_Suh}, for the time-invariant, symbol-extended interfering multiple access channels. The proposed upper bound is the first tight upper bound that is related to the idea of channel diversity \cite{Li_Jafarkhani_Jafar, Bresler_Tse_diversity} from the perspective of DoF. The achievability is obtained by the proposed modified scheme that systematically chooses the number of symbol extension and randomly chooses the precoding/combining vectors to impose the generic structure of precoding/combining matrices. We further extend our results to the time-invariant parallel MIMO interfering multiple access channels with independent subchannels.
There are several possible and important directions of the future work for this paper. For example, unlike the time-varying cases, even in the simple setting considered in this paper, the exact DoF characterization without the restriction of using linear pre- and post- processing is still an open question when the number of users in each cell is greater than the available channel diversity.
Other interesting directions are the extension to the asymmetric settings where the transmitters and receivers are equipped with different numbers of antennas and the extension to the settings with three of more cells.

\appendices

\newpage

\bibliography{Thesis}

\end{document}